\documentclass[epj,final]{svjour}
\usepackage{amssymb}
\usepackage{epsfig}
\begin{document}
\title{Hadron-Pair Photoproduction in Longitudinally Polarized Lepton-Nucleon
Collisions}
\author{C.\ Hendlmeier\inst{1} \and M.\ Stratmann\inst{1}\thanks{Address after 
August 15th, 2006: Radiation Laboratory, RIKEN, Wako, Saitama 351-0198, Japan}
 \and A.\ Sch{\"a}fer\inst{1}
}                     % Do not remove
\institute{\inst{1}
Institut f{\"u}r Theoretische Physik, Universit{\"a}t Regensburg,
D-93040 Regensburg, Germany}
\date{}
\abstract{We present a detailed phenomenological study of 
photoproduction of two hadrons, both with high transverse momentum,
in longitudinally polarized lepton-nucleon collisions. We
consistently include ``direct'' and ``resolved'' photon
contributions and examine the sensitivity of the relevant spin
asymmetries to the gluon polarization in the nucleon and to
the completely unknown parton content of circularly polarized
photons. Our results are relevant for the \textsc{Compass} and
\textsc{Hermes} fixed-target experiments as well as for a possible future
polarized lepton-proton collider like eRHIC at BNL. So far, all
studies are limited to the lowest order approximation of QCD. \PACS{
{13.88.+e}{}   \and
      {12.38.Bx}{}   \and
      {13.85.Ni} {}
     } % end of PACS codes
} %end of abstract

\maketitle
%%%%%%%%%%%%%%%%%%%%%%%%%%%%%%%%%%%%%
\section{Motivation and Introduction}
%%%%%%%%%%%%%%%%%%%%%%%%%%%%%%%%%%%%%
The fundamental question of how the spin of the proton is composed
of the spins and orbital angular momenta of its constituents, quarks
and gluons, still remains unanswered. Over the past 25 years, a
series of polarized deep-inelastic scattering (DIS) experiments has
revealed that the quark spins contribute remarkably little to the
nucleon spin \cite{ref:rith-review}. 
Measuring $\Delta g(x,\mu)$, the spin-dependent gluon
distribution in the nucleon, in an as large as possible range of
momentum fractions $x$, is the prime goal of all current experiments
with polarized beams and targets. In the light-cone gauge the first
moment of $\Delta g(x,\mu)$, i.e.~$\int_0^1 \Delta g(x,\mu) dx$, can be
interpreted as the gluon spin contribution to the nucleon spin
at a momentum scale $\mu$ \cite{ref:jaffe-manohar}. 
The missing piece, the orbital angular momenta of quarks and gluons,
might be accessible in exclusive processes, but precise measurements
are challenging and rather distant at this point.

The advent of the Relativistic Heavy Ion
Collider\linebreak (RHIC) at Brookhaven National Laboratory (BNL), has opened up
unequaled possibilities to access $\Delta g$ over a broad range in
$x$ in a variety of high-transverse momentum, ``high-$p_T$'', processes 
such as, for example, inclusive hadron or jet, prompt photon, 
and heavy flavor production \cite{ref:rhic-review}. 
In each case, the gluon density prominently contributes
through gluon-gluon fusion and quark-gluon scattering processes
already at the lowest order (LO) approximation of QCD.
Center-of-mass system (c.m.s.) energies of up to
$\sqrt{S}=500\,\mathrm{GeV}$ guarantee that the standard framework of
perturbative QCD (pQCD) can be used reliably to learn about all
aspects of helicity-dependent parton densities at RHIC.
A series of unpolarized ``benchmark'' measurements at
RHIC has nicely confirmed the applicability of 
pQCD methods \cite{ref:rhic-unpol} and are the foundation for similar, ongoing
measurements with polarization. 
First, very recent results from the
\textsc{Phenix} and \textsc{Star} collaborations at RHIC \cite{ref:rhic-pol}
indicate that large and positive gluon distributions are disfavored in the range
of momentum fractions $x$, $0.03\lesssim x\lesssim 0.2$, dominantly 
probed in these experiments. Future, more precise 
measurements will extend the range in $x$ and further close in on $\Delta g$.
 
The gluon polarization can be accessed also in low energy fixed-target 
experiments like {\sc Compass} \cite{ref:compass} at CERN and 
{\sc Hermes} \cite{ref:hermes} at DESY. 
Here one scatters a beam of longitudinally polarized leptons
off longitudinally polarized nucleons at c.m.s.\
energies of $\sqrt{S}\simeq 18\,\mathrm{GeV}$ and $\sqrt{S}\simeq
7.5\,\mathrm{GeV}$, respectively. 
Compared to RHIC, $\Delta g(x,\mu)$ is probed in a more limited $x$-range,
$0.1\lesssim x\lesssim 0.2$, but at smaller momentum scales $\mu$
which makes results complementary.
In particular, high-$p_T$ hadron
pairs, both in photoproduction and in deep-inelastic
electroproduction, have been identified to be the
most promising processes for a determination of $\Delta g$ at the
low energies available at fixed-target experiments
\cite{ref:twohadron}. 
First results for double-spin asymmetries are available from {\sc Hermes}
\cite{ref:hermes-2had} (for all photon virtualities),
{\sc Smc} \cite{ref:smc-2had} (electroproduction), and,
most recently, {\sc Compass} \cite{ref:compass-2had}
(photoproduction).  
These data are consistent with only moderate gluon polarizations
as we will discuss in detail later.

%%%%%%%%%%%%%%
% FIGURE 1
%%%%%%%%%%%%%%
\begin{figure*}[th]
\vspace*{0.cm}
\begin{center}
\begin{minipage}{7cm}
\epsfig{figure=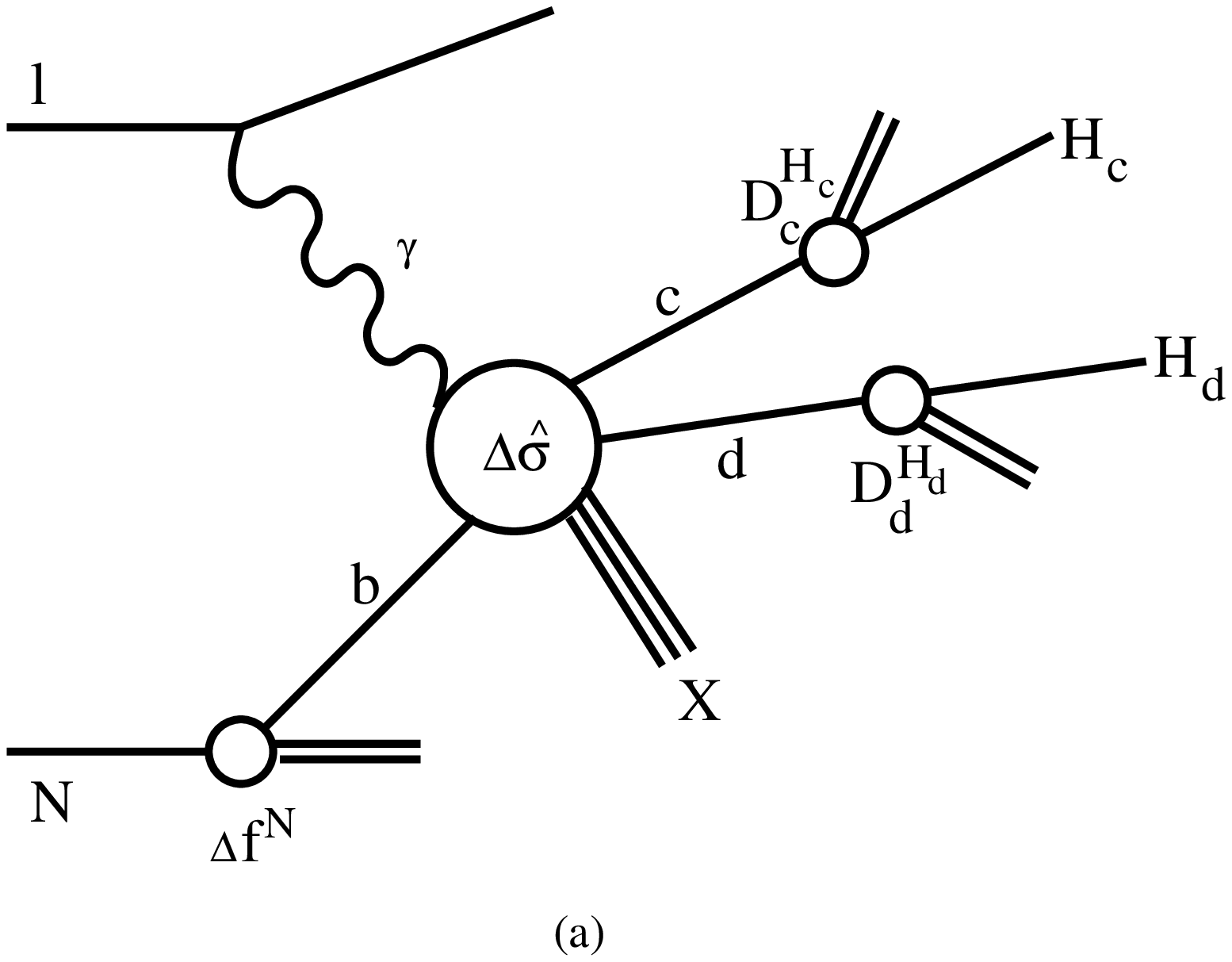,width=0.95\textwidth}
\end{minipage}
\begin{minipage}{7cm}
\epsfig{figure=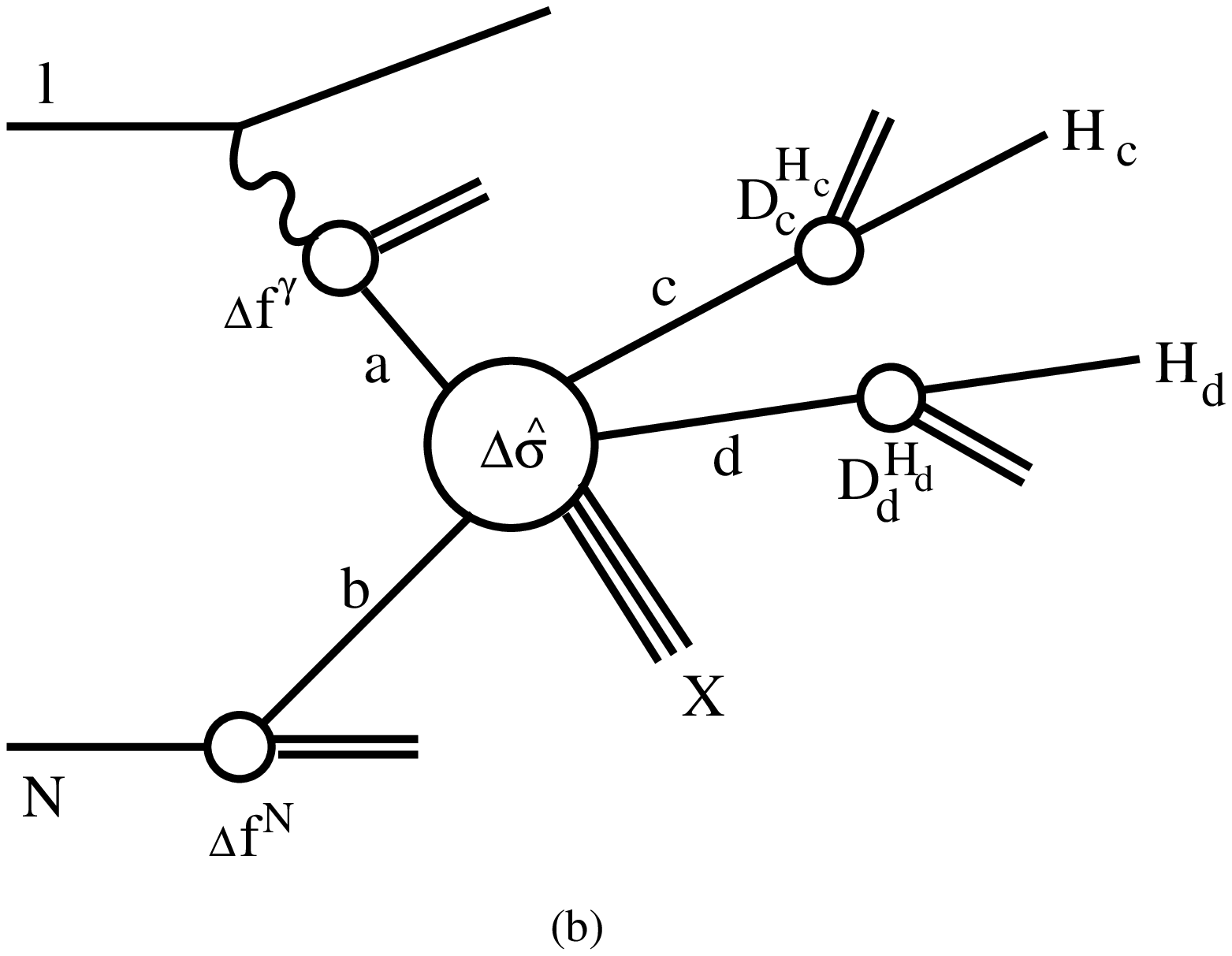,width=0.95\textwidth}
\end{minipage}
\end{center}
\vspace*{-0.25cm} \caption{\sf Generic direct {\bf (a)} and resolved 
{\bf (b)} photon contributions to the photoproduction of a pair of hadrons $H_c$
and $H_d$. \label{fig:cartoon}}
\end{figure*}
However, at the smaller c.m.s.\ energies of fixed-target experiments 
it is much less obvious that standard pQCD methods are applicable 
as straightforwardly as at collider energies to analyze data.
In fact, experimental results for high-$p_T$ processes in, e.g.,
hadron-hadron fixed-target scattering have been a
serious challenge  in the past for the standard factorized framework where the
perturbative series is truncated at a given fixed order in the
strong coupling and where possible power corrections are assumed to
be negligible \cite{ref:pp-trouble}.
It is therefore crucial to demonstrate {\em first} that standard pQCD 
methods can be used to learn about the
parton and/or spin content of nucleons in a given kinematical regime,
for instance, by analyzing unpolarized cross sections.
Otherwise conclusions about, e.g., the gluon polarization might be 
incorrect.

In this paper we present a detailed phenomenological study of 
photoproduction of hadron pairs at LO accuracy of QCD.
Quasi-real photons have the advantage of yielding much higher 
rates than deeply-inelastic electroproduction of hadrons. 
The price to pay is the more involved theoretical framework 
for photoproduction, where so-called ``direct'' and ``resolved'' 
photons contribute to the cross section as depicted in Figs.~1 (a) 
and (b), respectively. 
In (a) the photon simply interacts as an elementary
pointlike particle, whereas in the latter case 
the photon ``resolves'' into its parton
content prior to the hard QCD interaction, for instance, 
by fluctuating into a vector meson with the same quantum numbers.
Here, cross section estimates require knowledge of the
parton content of circularly polarized photons which is lacking
completely at the moment. 
We will demonstrate below, that this does not
seriously limit the usefulness of this process for the kinematical
region specific to both {\sc Compass} and {\sc Hermes}. 
Since the theoretical framework for hadron-pair
production is significantly more complex than for single-inclusive
cross sections, next-to-leading order (NLO) QCD corrections are
still not complete in the polarized case at the moment \cite{ref:upcoming}.
We note that there had been earlier studies of hadron-pair production 
\cite{ref:conto} in the wake of first experimental results 
from {\sc Hermes} \cite{ref:hermes-2had}. However, the 
phenomenological results presented in \cite{ref:conto} cannot be
compared easily with the recent (and upcoming) results from {\sc Compass}
\cite{ref:compass-2had} we are aiming at.

Investigations of interactions between polarized leptons and
nucleons will, hopefully, continue to play a vital role in spin
physics also in the future. A polarized lepton-nucleon collider such
as the eRHIC project at BNL \cite{ref:erhic}, which is currently 
under discussion, would be the next logical step. 
Besides the unique possibility to access $\Delta g(x,\mu)$ down to 
$x\simeq 10^{-3}$ in studies of scaling violations in DIS, 
photoproduction processes are particularly interesting also at 
collider energies \cite{ref:polhera}. As we shall show below, 
they will allow to probe for the first time different models 
for the parton content of circularly 
polarized photons.

The paper is organized as follows: in Sec.~2 we briefly recall the
theoretical framework for photoproduction of hadron pairs.
Section 3 is devoted to detailed phenomenological studies. Here we
mainly focus on the \textsc{Compass} and \textsc{Hermes}
experiments, where first data are already available. We 
present results for spin asymmetries and discuss their possible
sensitivity to $\Delta g$ but also focus on predictions for
unpolarized reference or ``benchmark'' cross sections which allow to
probe the validity of the pQCD framework at low c.m.s.\ energies and
transverse momenta $p_T$. We include all relevant
experimental cuts in our calculations. We close this section by
studying the prospects for hadron-pair photoproduction at eRHIC. We
briefly conclude in Sec.~4.

%%%%%%%%%%%%%%%%%%%%%%%%%%%%%
\section{Technical Framework}
%%%%%%%%%%%%%%%%%%%%%%%%%%%%%
We consider the spin-dependent inclusive photoproduction cross
section for the process
\begin{equation}
l(p_l)\,N(p_N) \to l^\prime (p_{l^{\prime}})\, H_c (p_c) \, H_d
(p_d)\, X\,\,\,,
\label{eq:photprod}
\end{equation}
where a longitudinally polarized lepton beam $l$ scatters off a
longitudinally polarized nucleon target $N$ producing two observed
hadrons $H_c$ and $H_d$ in the final state. The $p_i$ denote the
four-momenta of the particles. Both hadrons $H_c$ and $H_d$ are
assumed to have high transverse momenta $p_{T,c}$ and $p_{T,d}$,
respectively, ensuring large momentum transfer. Invoking the
factorization theorem \cite{ref:fact} we may then write the differential cross
section as a convolution of non-perturbative parton distribution and
fragmentation functions and partonic hard scattering cross sections,
\begin{eqnarray}
\label{eq:xsecdef}
\!\!\!\!\!\!\!\!d\Delta\sigma&\equiv&\frac{1}{2}[d\sigma_{++}-d\sigma_{+-}]=\\
&&\!\!\!\!\!\!\!\!\!\!\!\!\!\!\!\!\!\!\!\!\!\sum_{abcd} \int dx_a\, dx_b\, dz_c\, dz_d\,
\Delta f^l(x_a,\mu_f)\, \Delta f^N(x_b,\mu_f)\,\nonumber\\
&&\!\!\!\!\!\!\!\!\!\!\!\!\!\!\!\!\!\!\!\!\!\times \;
d\Delta\hat{\sigma}^{a b\to cdX'}(S,x_a,x_b,p_c/z_c,p_d/z_d,\mu_f,\mu_{f}',\mu_r)\nonumber\\
&&\!\!\!\!\!\!\!\!\!\!\!\!\!\!\!\!\!\!\!\!\!\times\;
D_c^{H_c}(z_c,\mu_{f}')\,
D_d^{H_d}(z_d,\mu_{f}')\,\,\,.
\label{eq:xsec}
\end{eqnarray}
In (\ref{eq:xsecdef}) the subscripts ``$++$'' and ``$+-$'' denote
the helicities of the colliding leptons and nucleons. $S$ is the
total c.m.s.\ energy squared available, i.e., $S=(p_l+p_N)^2$. The
sum in Eq.~(\ref{eq:xsec}) runs over all possible partonic channels
$a b\to c d$ with $d\Delta \hat{\sigma}^{a b\to c d}$ the associated
spin-dependent LO partonic hard scattering cross sections. The
latter can be calculated in pQCD order-by-order in the strong
coupling $\alpha_s(\mu_r)$, with $\mu_r$ denoting the
renormalization scale.

The $\Delta f^N(x_b,\mu_f)$ are the usual spin-dependent parton
distributions of the nucleon
\begin{equation}
\Delta
f^N(x_b,\mu_f)= f_{+}^N(x_b,\mu_f)-f_{-}^N(x_b,\mu_f)\,\,\,,
\label{eq:pdfdef}
\end{equation}
evolved to a factorization scale $\mu_f$, with $x_b$ the momentum
fraction of the nucleon carried by the parton $f$. The subscript $+$
$[-]$ in Eq.~(\ref{eq:pdfdef}) indicates that the parton's spin is
aligned [anti-aligned] to the spin of the parent nucleon. The other
non-perturbative functions $D_{c,d}^{H_{c,d}}(z_{c,d},\mu_{f}')$
describe the collinear fragmentation of the partons $c$ and $d$ into the
observed hadrons $H_c$ and $H_d$, respectively, with $z_{c,d}$ the
fraction of the parton's momentum carried by the produced hadron.
$\mu_{f}'$ denotes the final-state factorization scale which can be
different from $\mu_f$.

The experimentally measured cross section for (\ref{eq:photprod}) is
the sum of the so-called ``direct'' and ``resolved'' photon
contributions, cf.~Figs.~\ref{fig:cartoon} (a) and (b),
respectively,
\begin{equation}
d\Delta\sigma=d\Delta\sigma_{\mathrm{dir}}+d\Delta\sigma_{\mathrm{res}}\,\,\,.
\label{eq:sum}
\end{equation}
We shall note that neither $d\Delta\sigma_{\mathrm{dir}}$ nor
$d\Delta\sigma_{\mathrm{res}}$ are measurable individually.
Both, $d\Delta\sigma_{\mathrm{dir}}$ and
$d\Delta\sigma_{\mathrm{res}}$, can be cast into the form of
Eq.~(\ref{eq:xsec}) by defining the parton distribution functions
for a lepton, $\Delta f^l(x_a,\mu_f)$, appropriately. Most
generally, they can be written as convolutions,
\begin{equation}
\Delta f^l(x_a,\mu_f)=\int_{x_a}^1\frac{dy}{y}\Delta P_{\gamma
l}(y)\, \Delta
f^{\gamma}\left(x_{\gamma}=\frac{x_a}{y},\mu_f\right)\,\,\,,
\label{eq:flepton}
\end{equation}
with
\begin{eqnarray}
\Delta P_{\gamma l}(y)&=&\frac{\alpha_{em}}{2\pi}
\left[\frac{1-(1-y)^2}{y}\ln{\frac{Q_{max}^2(1-y)}{m_l^2y^2}}
\right.\nonumber\\
&+&\left.
2m_l^2y(\frac{1}{Q_{max}^2}-\frac{1-y}{m_l^2y^2})\right]
\label{eq:wwspectrum}
\end{eqnarray}
being the spin-dependent Weizs\"acker-Williams equivalent photon
spectrum \cite{ref:ww} that describes the collinear emission of a quasi-real
photon with momentum fraction $y$ and virtuality less than some
(small) upper limit $Q_{\mathrm{max}}$ off a lepton of mass $m_l$.
$Q_{\mathrm{max}}$ is determined by the experimental conditions.

The explicit form of the polarized photon structure function 
$\Delta f^{\gamma}\left(x_{\gamma},\mu\right)$ in Eq.(\ref{eq:flepton})
depends on the specifics of the interaction that the quasi-real
photon undergoes in the hard scattering with the nucleon. In the
``direct'' case, depicted in Fig.~1 (a), parton $a$ in
(\ref{eq:xsec}) has to be identified with an elementary photon and
hence $x_a$ with the momentum fraction $y$ of the photon w.r.t.\ the
parent lepton, i.e.,
\begin{equation}
\Delta f^{\gamma}(x_{\gamma},\mu)=\delta(1-x_{\gamma})
\end{equation}
in Eq.(\ref{eq:flepton}). If the photon resolves into its hadronic
structure before the hard scattering takes place, the $\Delta
f^{\gamma}$ in Eq.~(\ref{eq:flepton}) represent the parton densities
of a circularly polarized photon. The latter are defined in complete
analogy to the ones for a nucleon target in Eq.~(\ref{eq:pdfdef}).
Unlike hadronic parton distributions, photonic densities consist of
a perturbatively calculable ``pointlike'' contribution, which
dominates their behavior at large momentum fractions $x_{\gamma}$,
and a non-perturbative ``hadronic'' contribution dominating in the
low-to-mid $x_{\gamma}$ region. Nothing is known about the latter,
such that we have to invoke some model for it in our calculations
below. This will become important in the discussion of the numerical
results in the remainder of the paper. We will demonstrate that
measurements at low c.m.s.\ energies, i.e., at \textsc{Compass} and
\textsc{Hermes}, are to a large extent not affected by the actual
details of the model. At higher c.m.s.\ energies, like at a future
polarized $ep$ collider, one of the physics goals would be a first
determination of the partonic structure of circularly polarized
photons.

Finally, the experimentally relevant double-spin asymmetry $A_{LL}$
is defined as
\begin{equation}
A_{LL}\equiv \frac{d\Delta \sigma}{d\sigma}=
\frac{d\sigma_{++}-d\sigma_{+-}}{d\sigma_{++}+d\sigma_{+-}}\;.
\label{eq:all}
\end{equation}
The required spin-averaged cross section $d\sigma$
in Eq.~(\ref{eq:all}) is straightforwardly obtained from
Eqs.~(\ref{eq:xsec})-(\ref{eq:wwspectrum}) by replacing all
polarized quantities by their appropriate unpolarized counterparts.

%%%%%%%%%%%%%%%%%%%%%%%%%%%%%%%%%%%%%%%
\section{Phenomenological Applications}
%%%%%%%%%%%%%%%%%%%%%%%%%%%%%%%%%%%%%%%
%
%%%%%%%%%%%%%%%
% FIGURE 2
%%%%%%%%%%%%%%%
\begin{figure*}[tp]
\vspace*{-0.8cm}
\begin{center}
\begin{minipage}{7.5cm}
\epsfig{figure=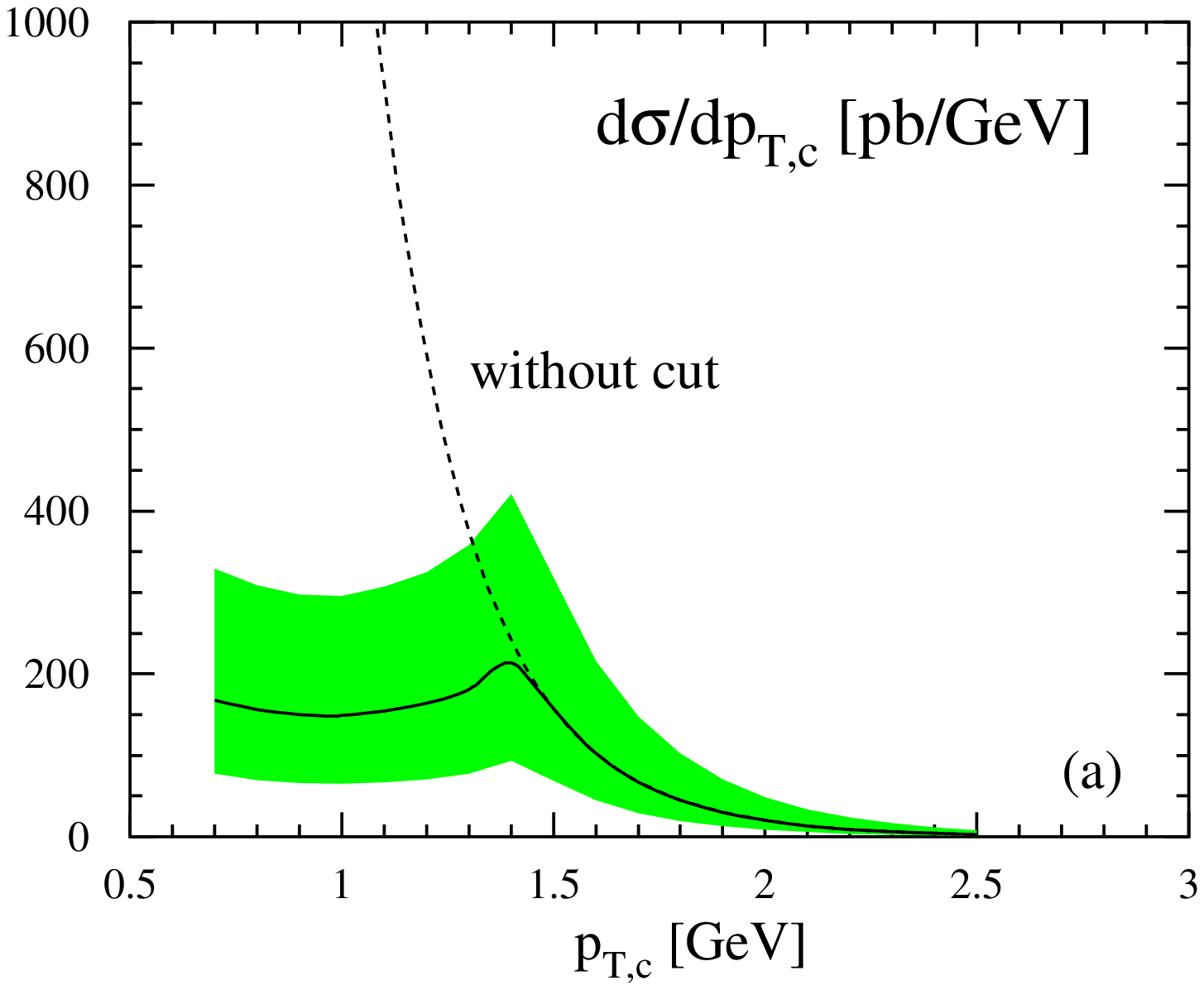,width=1.2\textwidth}
\end{minipage}
\hspace*{1.0cm}
\begin{minipage}{7.5cm}
\epsfig{figure=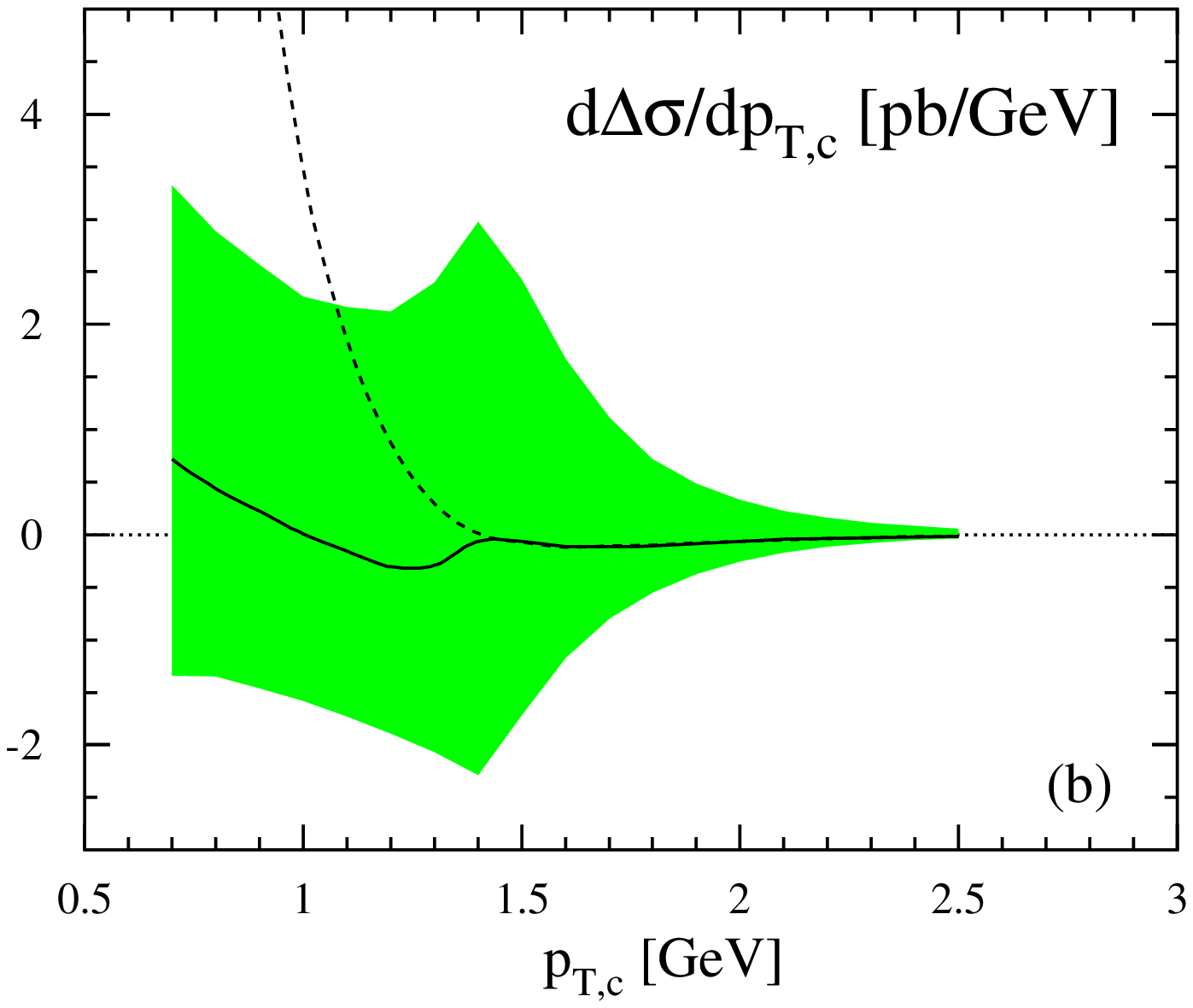,width=1.2\textwidth}
\end{minipage}
\end{center}
\vspace*{-0.5cm}
\caption{\sf Scale dependence of the unpolarized {\bf (a)} and 
polarized {\bf (b)} photoproduction cross section for two charged
hadrons at LO for \textsc{Compass} with $\theta_{\max}=70\,\mathrm{mrad}$. 
The scales in Eq.~(\ref{eq:xsec}) are varied 
simultaneously in the range $(p_{T,c}^2+p_{T,d}^2)/4\leq \mu^2 \leq
4(p_{T,c}^2+p_{T,d}^2)$ as indicated by the shaded bands.
The solid lines correspond to the default choice where
$\mu^2=p_{T,c}^2+p_{T,d}^2$. 
The dotted lines show the impact of lifting the
experimental cut $p_{T,c}^2+p_{T,d}^2>2.5\;\mathrm{GeV}^2$.
\hfill\mbox{}}
\label{fig:scadepcomp}
\end{figure*}
In our phenomenological studies based on the framework laid out in 
Eqs.~(\ref{eq:xsec})-(\ref{eq:wwspectrum})
we concentrate on the production of pairs of charged hadrons made of
light quark flavors. In fact, we sum over pions, kaons, and (anti-)protons and use the
fragmentation functions of KKP \cite{ref:kkp} throughout to model 
hadronization. 
All our results will be differential in the transverse
momentum $p_{T,c}$ of hadron $H_c$ and integrated over all 
kinematically and experimentally allowed transverse momenta
$p_{T,d}$ of hadron $H_d$ and pseudo-rapi\-di\-ties $\eta_{c,d}$ 
unless stated otherwise. The pseudo-rapidities of the hadrons 
are measured w.r.t.\ the direction of the incident lepton beam.

For the unpolarized parton densities of the nucleon
and photon we adopt the LO CTEQ6L \cite{ref:cteq} and GRV
\cite{ref:grvphoton} sets, respectively.
To study the sensitivity to the
unknown gluon polarization of the nucleon we use four different sets of
spin-dependent parton distributions emerging from the GRSV analysis \cite{ref:grsv}.
These sets span a rather large range of gluon densities $\Delta g$
all very much consistent with present DIS data. Apart from our default ``standard'' 
set of GRSV with a moderately large, positive $\Delta g$, the three other sets
``$\Delta g=g$ input'', ``$\Delta g=0$ input'', and ``$\Delta g=-g$ input''
are characterized by a large positive, a vanishing, and a large negative gluon
polarization, respectively, at the input scale of the evolution.

The unknown parton densities of 
circularly polarized photons are estimated with the help of
two extreme models \cite{ref:polphoton} based on 
maximal, $\Delta f^{\gamma}(x,\mu_0)= f^{\gamma}(x,\mu_0)$,
or minimal, $\Delta f^{\gamma}(x,\mu_0)=0$,
saturation of the positivity bound at the starting scale $\mu_0$ 
for the evolution to scales $\mu>\mu_0$.
Both models result in very different parton distributions $\Delta f^\gamma$ at
small-to-medium $x_\gamma$ while they almost coincide as $x_\gamma\to 1$
due to the dominance of the perturbatively calculable 
``pointlike'' contribution in this region. 
The use of the ``maximal'' set will be implicitly understood
except when we study the sensitivity of the photoproduction cross sections
and spin asymmetries to the details of the
non-perturbative hadronic input to the evolution of $\Delta f^\gamma$.

Unless stated otherwise, all factorization and renormalization scales,
$\mu_f$, $\mu_{f}'$, and $\mu_r$, in Eq.~(\ref{eq:xsec}) are set equal to
$\mu^2\equiv\mu_r^2=\mu_f^2=\mu_f'^2= p_{T,c}^2+p_{T,d}^2$.
%
%%%%%%%%%%%%%%%%%%%%%%%%%%%%%%%%%%%%%%%%%%%%%%
\subsection{Two-Hadron Production at COMPASS}
%%%%%%%%%%%%%%%%%%%%%%%%%%%%%%%%%%%%%%%%%%%%%%
With the present setup, the \textsc{Compass} experiment
scatters polarized muons with a beam energy of
$E_{\mathrm{\mu}}=160\;\mathrm{GeV}$ off the deuterons in a polarized
${}^6\mathrm{Li}\mathrm{D}$ solid-state target corresponding to a
c.m.s.\ energy of $\sqrt{S}\simeq 18\;\mathrm{GeV}$.
On average the beam polarization is ${\cal{P}}_{\mu}\simeq 76\%$, and
about ${\cal{F}}_d\simeq 50\%$ of the deuterons can be polarized with an average
polarization of ${\cal{P}}_d\simeq 50\%$ \cite{ref:compass}.

Hadrons can be detected if their scattering angle is less than
$\theta_{\mathrm{max}}=70\;\mathrm{mrad}$ in the laboratory frame.
This acceptance was recently upgraded to $\theta_{\mathrm{max}}=180\;\mathrm{mrad}$
for all future runs.
In the event selection for a ``high-$p_T$'' sample, the charged hadrons
have to pass further cuts \cite{ref:compass-2had}: the invariant mass $m(H_c,H_d)$ 
of the two produced hadrons has to be larger than $1.5\;\mathrm{GeV}$ and the
sum of the transverse momenta squared must exceed 
$p_{T,c}^2+p_{T,d}^2>2.5\;\mathrm{GeV}^2$ with both $p_{T,c}$ and 
$p_{T,d}$ larger than $0.7\,\mathrm{GeV}$.
In addition, the fractions $z_{c,d}$ of the parent parton's momenta
carried by the detected hadrons $H_{c,d}$ are chosen to be $z_{c,d}>0.1$.

The maximal virtuality of the quasi-real photons in 
Eq.~(\ref{eq:wwspectrum}) is taken to be $Q^2_{\mathrm{max}}=0.5\;\mathrm{GeV}^2$. 
The fraction $y$ of the lepton's momentum taken by the photon is restricted 
to be in the range $0.1\leq y\leq 0.9$.
We note that the often omitted non-logarithmic pieces in Eq.~(\ref{eq:wwspectrum})
result in a small but non-negligible contribution for muons.

Figure \ref{fig:scadepcomp} shows the dependence of both the unpolarized and
polarized LO photoproduction cross section (\ref{eq:xsec}) 
on the unphysical factorization/renormalization scales varied in the range
$(p_{T,c}^2+p_{T,d}^2)/4\leq\mu^2\leq 4(p_{T,c}^2+p_{T,d}^2)$.
Both cross sections exhibit a very large scale dependence which is, however, 
not uncommon for LO estimates.
Sets of polarized parton densities with a moderate gluon polarization 
like GRSV ``standard'' result in an almost vanishing cross section
as the two ``direct'' subprocesses, photon-gluon-fusion and 
QCD-Compton scattering, cancel each other almost entirely, see also 
Fig.~\ref{fig:ratiocomppol} and the
discussions below. Even the sign of the polarized cross 
section cannot be determined within the scale uncertainty here. 
As always, the computation of the relevant NLO QCD corrections is a mandatory
task as theoretical uncertainties associated with the residual scale dependence
are due to the truncation of the perturbative series at a given order and
are expected to decrease significantly beyond the LO approximation.
Such a calculation for hadron-pair production is a formidable task 
and still not complete at present \cite{ref:upcoming}.

%
%%%%%%%%%%%%%%%
%%% FIGURE 3
%%%%%%%%%%%%%%%
\begin{figure}[tp]
\begin{center}
\vspace*{-.8cm}
\includegraphics[width=0.49\textwidth,clip=]{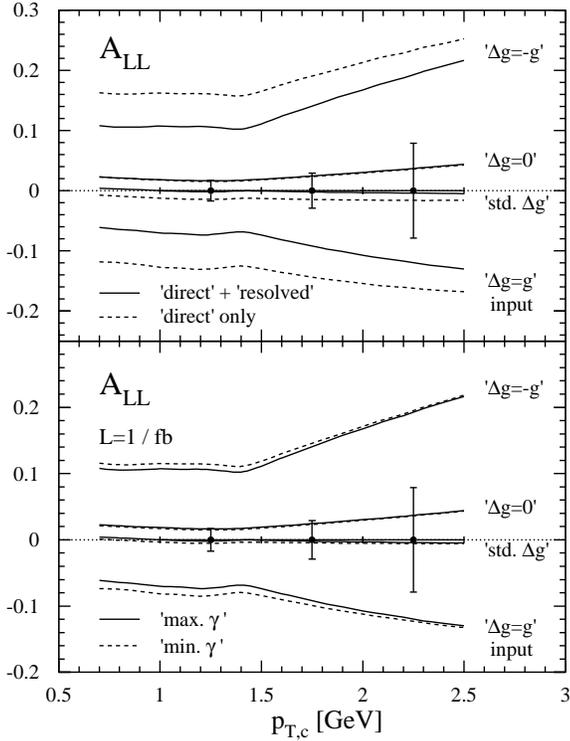}
\caption{\sf Double-spin asymmetry $A_{LL}$, Eq.~(\ref{eq:all}), at LO for
different gluon polarizations in the nucleon (see text). 
The upper panel shows the spin asymmetry with (solid lines) and
without (dashed lines) including the ``resolved'' contribution.
In the lower panel the dependence of $A_{LL}$ on the two extreme
photon scenarios, ``maximal'' (solid lines) and ``minimal''
saturation (dashed lines), is shown. 
The ``error bars'' indicate the estimated
statistical accuracy for such a measurement at \textsc{Compass}
in certain bins of $p_{T,c}$ based on an integrated luminosity 
of $1\,\mathrm{fb}^{-1}$. \hfill\mbox{}}
\label{fig:asycomp}
\end{center}
\end{figure}
From a recent calculation of the NLO QCD corrections to
spin-dependent, {\em single-inclusive} photoproduction of high-$p_T$ hadrons
\cite{ref:onehadron} we know though that theoretical scale
uncertainties are only marginally reduced, if at all, at the NLO level 
at c.m.s.\ energies relevant for {\sc Compass} and {\sc Hermes}.
This is in sharp contrast to what one generally expects and what indeed
happens at collider energies, see, e.g., \cite{ref:pionpp,ref:pionerhic}.
In addition, the NLO corrections in \cite{ref:onehadron} turned out to be
sizable and different for the polarized and unpolarized cross sections thus 
affecting also the spin asymmetries defined in Eq.~(\ref{eq:all}).
Clearly, this underpins the delicacy of perturbative calculations in the
fixed-target regime, i.e., for small c.m.s.\ energies and
transverse momenta of only a few GeV.
Before drawing any conclusions about the hadronic spin structure,
in particular $\Delta g$, from measurements of spin asymmetries $A_{LL}$
one has to demonstrate the applicability of pQCD methods first. 
For this purpose, an important ``benchmark'' would be the comparison of 
the relevant unpolarized cross sections with theoretical expectations, e.g.,
given in Fig.~\ref{fig:scadepcomp} (a).
Unfortunately, such kind of information is still lacking from 
{\sc Compass} for the time being. 
Let us stress, that a possible discrepancy between experiment 
and theory at moderate c.m.s.\ energies
would not necessarily imply that standard pQCD methods are beyond remedy.
It would only call for further improvements by resumming
large terms in the perturbative series, for instance,
threshold logarithms, to all orders in the strong coupling.
This is known to lead often to a much improved agreement between data and pQCD
calculations, see, e.g., \cite{ref:resum}.

The dotted lines in Fig.~\ref{fig:scadepcomp} correspond to the 
unpolarized (a) and polarized (b) cross sections computed without imposing
the experimental cut on $p_{T,c}^2+p_{T,d}^2>2.5\;\mathrm{GeV}^2$.
This cut, which ensures hard scattering, is responsible for the cusp 
observed at around $p_{T,c}\simeq 1.4\,\mathrm{GeV}$ and for the significant 
reduction of the cross section for smaller $p_{T,c}$.

Figure \ref{fig:asycomp} shows our expectations for the 
double-spin asymmetry $A_{LL}$, Eq.~(\ref{eq:all}), at LO
based on the cross sections shown in Fig.~\ref{fig:scadepcomp} for the
default choice of scales.
Apart from the ``standard'' set of GRSV polarized parton densities \cite{ref:grsv}, 
we also use the three other sets, introduced at the beginning of Sec.~3,
with very different assumptions about the gluon polarization.
In the upper panel of Fig.~\ref{fig:asycomp} we study the importance of
the ``resolved'' photon contribution to the photoproduction cross section
(\ref{eq:sum}). By comparing the experimentally relevant spin asymmetry
for the sum of ``direct'' and ``resolved'' contributions (solid lines)
with $A_{LL}$ computed for the ``direct'' part alone (dashed lines) one can infer
that irrespective of the chosen $\Delta g$  the ``resolved'' part
is non-negligible. It leads to a significant shift in the absolute value
of the spin asymmetry and neglecting it in the analysis would clearly
lead to wrong conclusions about $\Delta g$.

The impact of the unknown, non-perturbative parton content of the circularly
polarized photon on $A_{LL}$ is examined in the lower panel 
of Fig.~\ref{fig:asycomp} by making use of the two extreme
models \cite{ref:polphoton} also introduced at the beginning of Sec.~3.
As can be seen, the actual choice of the model barely affects the results
for the spin asymmetry. Any difference between the two results diminishes 
further towards larger $p_{T,c}$.
This can be readily understood as the photonic parton densities
are probed on average at medium-to-large momentum fractions $x_{\gamma}$.
In this region the partonic content of the photon is dominated by the
``pointlike'' contribution which is independent of the details of the
unknown non-perturbative input \cite{ref:polphoton}. 
Certainly, this finding somewhat simplifies the theoretical analysis of
the spin asymmetry in terms of $\Delta g$.

Figure \ref{fig:asycomp} also demonstrates that the double-spin asymmetry $A_{LL}$ 
is sensitive to different model assumptions for the gluon polarization. 
A large and positive gluon polarization yields a sizable negative asymmetry 
whereas a large and negative $\Delta g$ leads to a positive asymmetry. 
For the ``standard'' set of GRSV or when $\Delta g=0$ is imposed
at the input scale of the evolution we find asymmetries close to zero.
To judge whether a measurement of $A_{LL}$ can be actually turned into a 
constraint on $\Delta g$ we estimate the expected statistical 
accuracy $\delta A_{LL}$ for {\sc Compass}
in certain bins of $p_{T,c}$, calculated from
\begin{equation}
\delta A_{LL} \simeq \frac{1}{{\cal{P}}_{\mu}{\cal{P}}_d{\cal{F}}_d}
\frac{1}{\sqrt{\sigma_{bin}{\cal{L}}}}\;\;.
\label{eq:all-error}
\end{equation}
Here, $\sigma_{bin}$ denotes the unpolarized cross section integrated over the
bin considered and ${\cal{L}}$ the integrated luminosity for which we assume
${\cal{L}}=1\;\mathrm{fb}^{-1}$. All other quantities are as specified at the
beginning of Sec.~3.
Clearly, the region of $1<p_{T,c}<2\;\mathrm{GeV}$ is the most promising one
to obtain information about the gluon polarization. 
At higher $p_{T,c}$'s the achievable statistical precision deteriorates
as the cross section drops steeply with $p_{T,c}$.

First experimental results \cite{ref:compass-2had} find that the 
spin asymmetry (integrated also over $p_{T,c}$ for the time being) 
is close to zero. From this measurement a value of 
$\Delta g/g=0.024\pm 0.089(\mathrm{stat.})\pm 0.057(\mathrm{sys.})$
at $x=0.095+0.08\,(-0.04)$ and scale $\mu^2=3\,\mathrm{GeV^2}$
was extracted with the help of ``purities'', i.e., a ``signal-to-background''
separation based on Monte-Carlo simulations \cite{ref:compass-2had}.
Compared to our theoretical expectations in Fig.~\ref{fig:asycomp}
the {\sc Compass} result of $A_{LL}\simeq 0$ \cite{ref:compass-2had} is 
consistent with  a moderate gluon polarization like in the GRSV ``standard'' set.
However, as discussed above, one should take this result with a grain of salt unless
an unpolarized reference cross section becomes available from {\sc Compass}
to verify the applicability of pQCD methods. 
In particular, one has to exclude that the observed 
smallness of the spin asymmetry is due 
to the presence of large non-perturbative effects. If these are  
spin-independent they would naturally lead to $A_{LL}\simeq 0$ irrespective of
$\Delta g$.
%%%%%%%%%%%%%%%
%%% FIGURE 4
%%%%%%%%%%%%%%%
\begin{figure}[tp]
\begin{center}
\vspace*{-.8cm}
\includegraphics[width=0.49\textwidth,clip=]{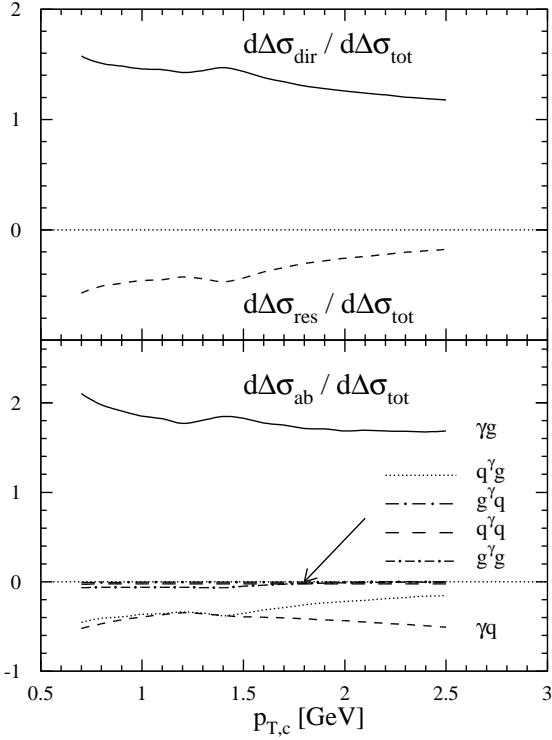}
\caption{\sf The upper panel shows the
``direct'' and ``resolved'' contribution to the experimentally
relevant polarized cross section. The lower panel
shows the fractional contributions of the different partonic LO channels 
$a+b\to c+d$ to the photoproduction cross section.
\hfill\mbox{}}
\label{fig:ratiocomppol}
\vspace*{-0.2cm}
\end{center}
\end{figure}

Next we turn to a closer analysis of how the results in Figures
\ref{fig:scadepcomp} and \ref{fig:asycomp} can be understood. 
To this end we study the different contributions to the polarized 
photoproduction cross section separately as illustrated in 
Figure \ref{fig:ratiocomppol}. The upper and lower panel show the fractional
contributions of $d\Delta\sigma_{\mathrm{dir}}$ and  $d\Delta\sigma_{\mathrm{res}}$
and of the different partonic LO channels $a+b\to c+d$, respectively.
Here we use the maximal positive gluon polarization of GRSV
with $\Delta g=g$ at the input scale, 
for which the cancellation between the photon-gluon fusion and QCD Compton 
subprocesses mentioned above is less relevant.
For our ``default'' gluon polarization used in Fig.~\ref{fig:scadepcomp} (b)
the polarized cross section has a node at some $p_{T,c}$ such
that ratios are difficult to visualize.
From Fig.~\ref{fig:ratiocomppol} one infers that the ``direct'' part dominates 
in absolute value in the entire $p_{T,c}$-range shown and that the ``resolved''
and ``direct'' contributions have opposite signs. 
Turning to the individual subprocesses in the lower panel of 
Fig.~\ref{fig:ratiocomppol} we note that the QCD Compton $\gamma q$-channel 
always gives a positive contribution to the cross section\footnote{$d\Delta\sigma_{\mathrm{tot}}<0$
for the large and positive gluon polarization used in Fig.~\ref{fig:ratiocomppol} such that
$d\Delta\sigma_{\gamma q}/d\Delta\sigma_{\mathrm{tot}}<0$.} whereas the sign of the 
photon-gluon fusion channel is anti-correlated with the sign of $\Delta g$
and its relevance scales with the magnitude of the unknown gluon polarization.
This leads to a partial cancellation between the two ``direct'' channels for
positive gluon polarizations which explains the smallness of the 
polarized cross section for the  GRSV ``standard'' gluon observed 
in Fig.~\ref{fig:scadepcomp} (b).
Of the ``resolved'' processes
only the scattering of a quark with large momentum fraction in the photon 
off a gluon in the nucleon makes a significant contribution, 
other channels are negligible.
We note that in the unpolarized case (not shown here)
all contributions to the cross section are positive with
the ``direct'' part accounting for 80 percent or more.
%
%%%%%%%%%%%%%%%
%%% FIGURE 5
%%%%%%%%%%%%%%%
\begin{figure}[t]
\begin{center}
\vspace*{-.8cm}
\includegraphics[width=0.49\textwidth,clip=]{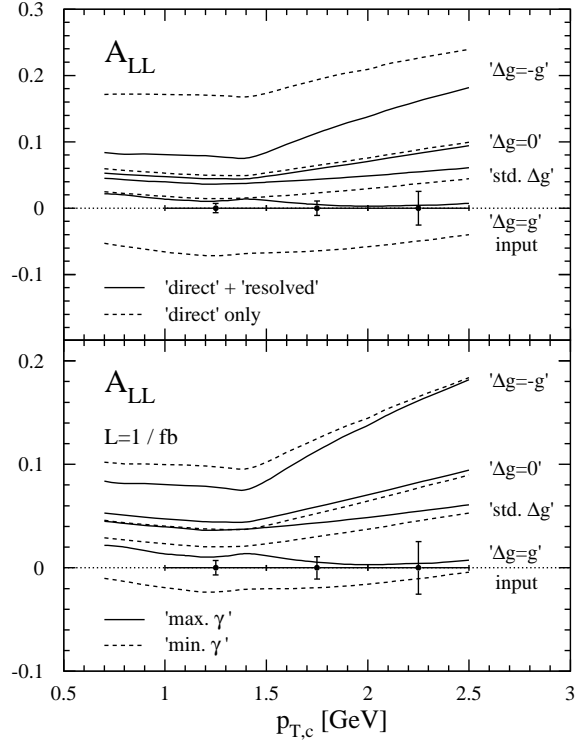}
\caption{\sf As in Figure \ref{fig:asycomp} but now for the 
improved angular acceptance of 
$\theta_{\mathrm{max}}=180\;\mathrm{mrad}$.\hfill\mbox{}}
\label{fig:asycomp180}
\vspace*{-0.2cm}
\end{center}
\end{figure}

We also wish to comment on the momentum fractions $x_b$ predominantly probed 
in the nucleon in a measurement of hadron-pair photoproduction at {\sc Compass}.
For the unpolarized cross section this can be easily specified by looking at the
distribution in $x_b$ for a given bin of $p_{T,c}$. For example, for
$p_{T,c}$ around $1\,\mathrm{GeV}$ we find $\langle x_b \rangle = 0.12\pm 0.05$
which is consistent with the $x$-range estimated by 
{\sc Compass} \cite{ref:compass-2had}.
However, similar estimates for the polarized cross section and hence for $\Delta g/g$ are
impossible without knowing $\Delta g$ beforehand
as both the polarized cross sections and
the helicity parton distributions are not positive definite. The relevance of 
contributions of opposite sign to the cross section strongly depends on
the gluon polarization and completely obscures the meaning of an 
averaged $\langle x_b \rangle$ here.
This issue can be only consistently resolved in a future global analysis of 
polarized parton densities.
%%%%%%%%%%%%%%%
% FIGURE 6
%%%%%%%%%%%%%%%
\begin{figure*}[tp]
\vspace*{-0.8cm}
\begin{center}
\begin{minipage}{7.4cm}
\epsfig{figure=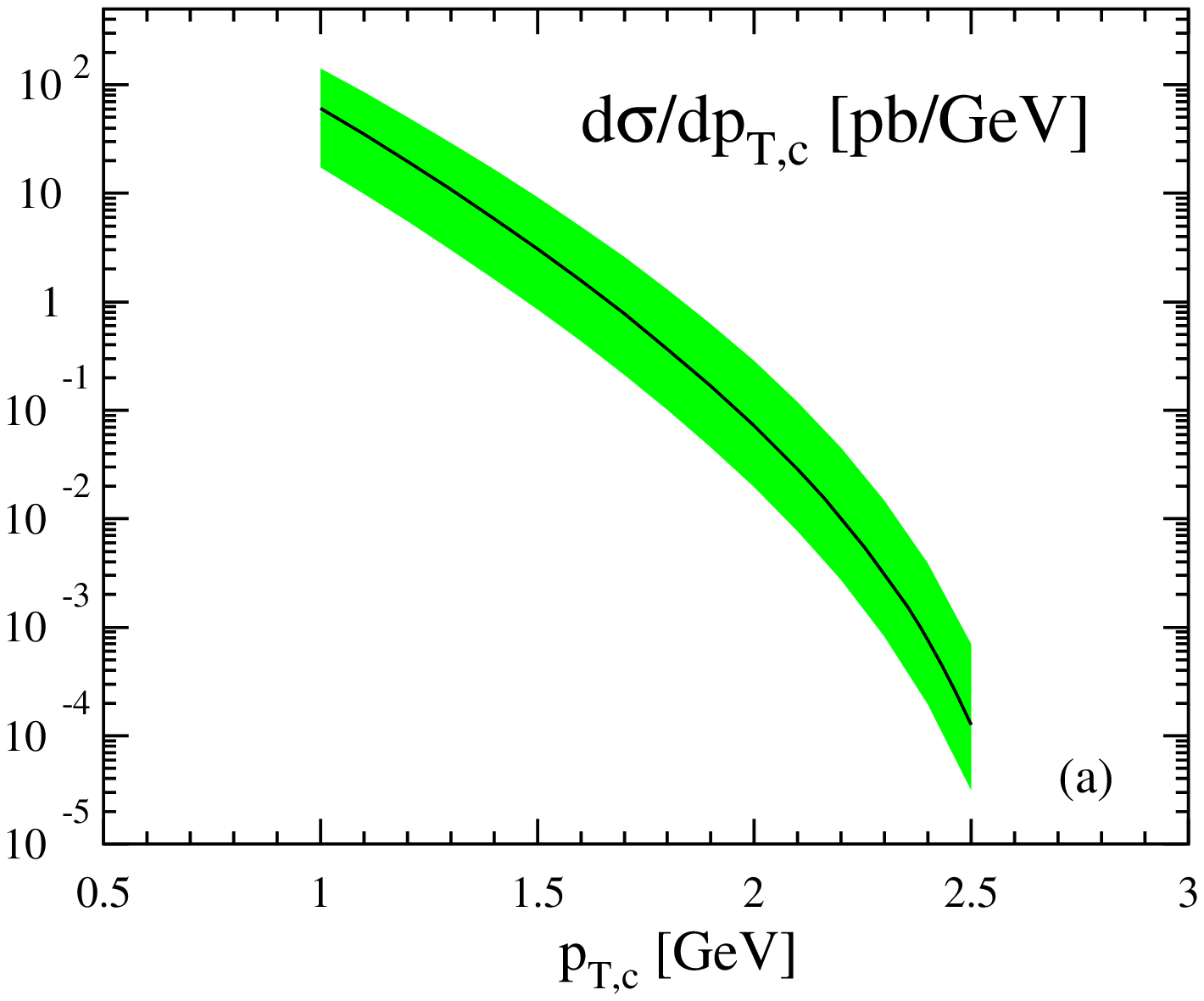,width=1.2\textwidth}
\end{minipage}
\hspace*{1.0cm}
\begin{minipage}{7.4cm}
\epsfig{figure=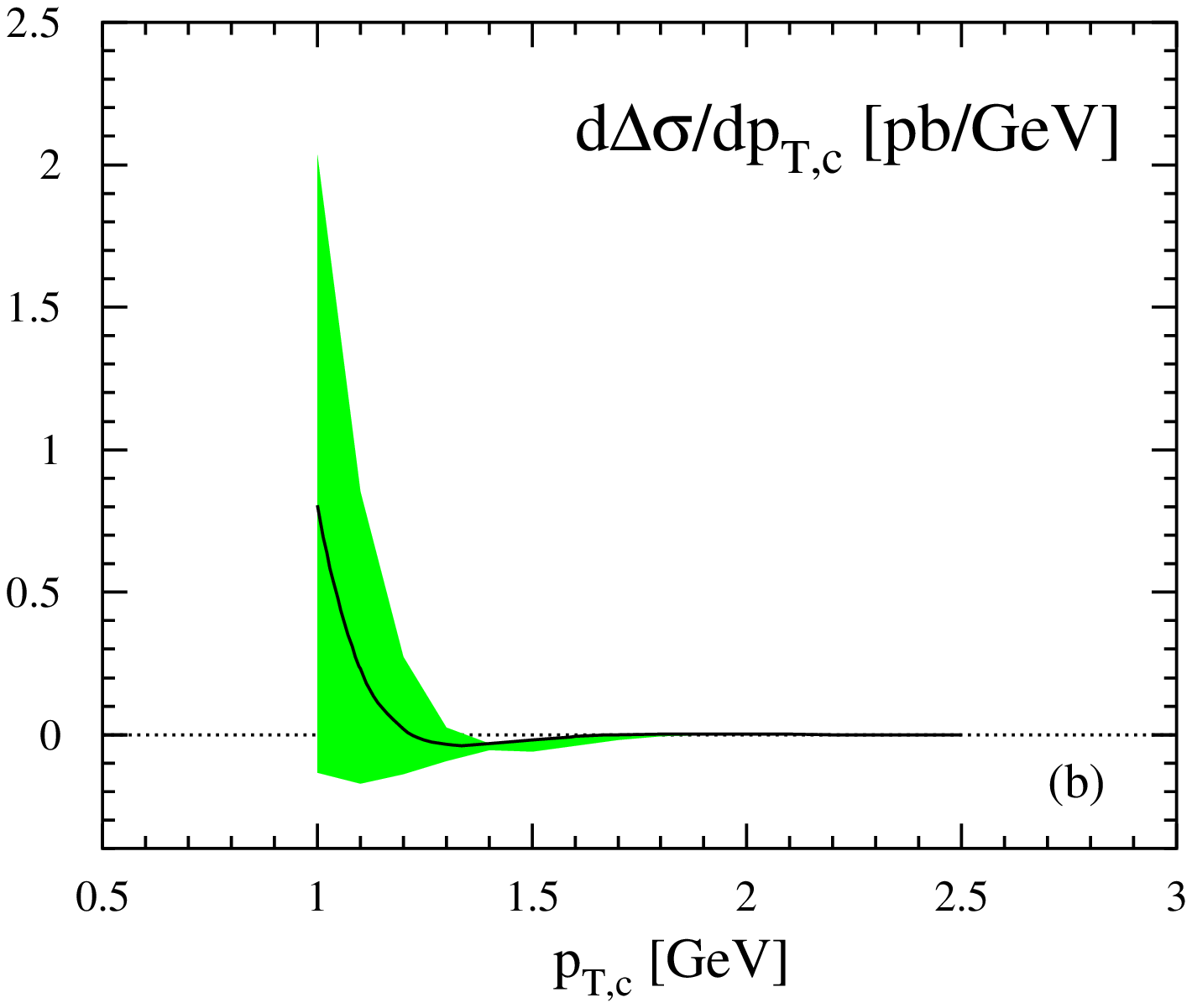,width=1.2\textwidth}
\end{minipage}
\end{center}
\vspace*{-0.5cm}
\caption{\sf Scale dependence of the unpolarized {\bf (a)} and polarized 
{\bf (b)} photoproduction cross section for two charged hadrons in LO 
for \textsc{Hermes}. The scales in Eq.~(\ref{eq:xsec}) 
are varied simultaneously in the range $(p_{T,c}^2+p_{T,d}^2)/4\leq\mu^2\leq
4(p_{T,c}^2+p_{T,d}^2)$ as indicated by the shaded bands. 
The solid lines correspond to the default choice where
$\mu^2=p_{T,c}^2+p_{T,d}^2$.\hfill\mbox{}}
\label{fig:scadepherm}
\vspace*{-0.2cm}
\end{figure*}

Next it is interesting to check whether the upgrade of the angular acceptance
of the \textsc{Compass} experiment to $\theta_{\mathrm{max}}=180\;\mathrm{mrad}$
will enhance their sensitivity to $\Delta g$.
In Fig.~\ref{fig:asycomp180} we present our expectations for the double-spin asymmetry 
$A_{LL}$ as a function of the transverse momentum $p_{T,c}$ of one of the hadrons. 
Except for $\theta_{\mathrm{max}}=180\;\mathrm{mrad}$ all other settings and
cuts are the same as in Fig.~\ref{fig:asycomp}.
The statistical precision for such a measurement is  
again estimated with the help of Eq.~(\ref{eq:all-error})
for an integrated luminosity of $1\,{\mathrm{fb}}^{-1}$.
From the upper panel of Fig.~\ref{fig:asycomp180} one infers that the 
``resolved'' contribution modifies the asymmetry even more significantly
than for $\theta_{\mathrm{max}}=70\;\mathrm{mrad}$.
Even worse, there is now also a strong dependence on the model used to
describe the parton content of the circularly polarized photon as can be seen 
in the lower panel of Fig.~\ref{fig:asycomp180}.
This can be readily understood by noticing that due to the larger angular coverage
one now probes the partonic structure of the photon also at momentum
fractions $x_{\gamma}$ where the details of the unknown non-perturbative input do matter.
Only in the high-$p_{T,c}$-region, where $x_{\gamma}\to 1$, the dependence
on the model for $\Delta f^{\gamma}$ becomes small. 
Clearly, a viable strategy would be to analyze data with different cuts on
$\theta_{\max}$ or for bins in $\theta$ 
and to learn as much as possible about $\Delta g$ first. 
Data up to $\theta_{\mathrm{max}}=180\;\mathrm{mrad}$ might then be used for
studying the non-perturbative structure of circularly polarized photons.
Needless to mention again, that the validity of the pQCD framework 
for two-hadron production at {\sc Compass} has to be confirmed prior to
studies of $A_{LL}$.

%%%%%%%%%%%%%%%%%%%%%%%%%%%%%%%%%%%%%%%%%%%%%
\subsection{Two-Hadron Production at HERMES}
%%%%%%%%%%%%%%%%%%%%%%%%%%%%%%%%%%%%%%%%%%%%%
At the \textsc{Hermes} experiment at DESY 
longitudinally polarized electrons/positrons with a beam energy
of $E_e\simeq 27.5\; \mathrm{GeV}$ were scattered off both, a polarized deuterium 
or a polarized hydrogen gas target. 
The available c.m.s.\ energy of about $\sqrt{S}\simeq 7.5$~GeV
is lower than at {\sc Compass} which even further limits the range
of accessible transverse momenta.
On average the lepton beam polarization is ${\cal{P}}_{e}\simeq 53\%$.
For the polarization of the gas target we take ${\cal{P}}_d\approx{\cal{P}}_p\simeq 85\%$,
and, contrary to a solid-state target, there is no dilution,
i.e., ${\cal{F}}_p={\cal{F}}_d=1$.

We concentrate on phenomenological studies for a polarized deuterium target
in line with the data sample with the highest statistics in the
{\sc Hermes} spin physics program which came to an end recently.
The \textsc{Hermes} experiment has an angular acceptance of
$40 \;\mathrm{mrad} \leq \theta_{\mathrm{lab}}\leq 220 \;\mathrm{mrad}$
for hadrons. For all our numerical studies we demand a transverse
momentum of at least $1\,\mathrm{GeV}$ for both detected hadrons $H_{c,d}$.
We choose a maximal photon virtuality of $Q^2_{\mathrm{max}}=0.1\;\mathrm{GeV}^2$
in Eq.~(\ref{eq:wwspectrum}) and restrict $y$ to $0.2\leq y\leq 0.9$.
The fractions of the parent parton's momenta carried
by the produced hadrons are
$z_{c,d}\geq 0.1$. Again, all scales in Eq.~(\ref{eq:xsec}) are set equal to
$\mu^2=p_{T,c}^2+p_{T,d}^2$ unless  stated otherwise.

Figure \ref{fig:scadepherm} shows the dependence of both the unpolarized and
polarized LO photoproduction cross section, Eq.~(\ref{eq:xsec}), 
on the unphysical factorization/renormalization scales varied in the range
$(p_{T,c}^2+p_{T,d}^2)/4\leq\mu^2\leq 4(p_{T,c}^2+p_{T,d}^2)$.
Not unexpectedly, due to the smaller c.m.s.~energy of the \textsc{Hermes} experiment, 
the scale dependence is even larger than for \textsc{Compass}, 
cf.\ Fig.~\ref{fig:scadepcomp}.
All remarks about potential problems with the applicability of
perturbative methods at fixed-target energies and the need for unpolarized
``benchmark'' cross sections also apply here.

%%%%%%%%%%%%%%%
%%% FIGURE 7
%%%%%%%%%%%%%%%
\begin{figure}
\begin{center}
\vspace*{-.8cm}
\includegraphics[width=0.49\textwidth,clip=]{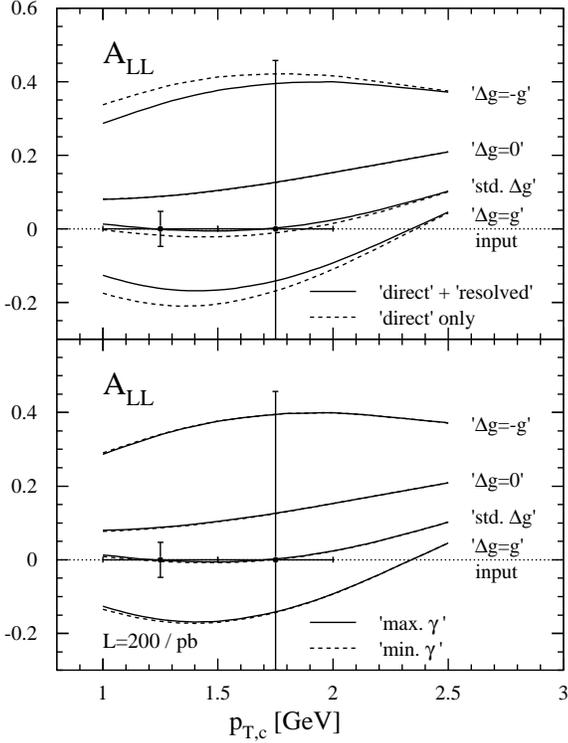}
\caption{\sf As in Fig.~\ref{fig:asycomp} but know for \textsc{Hermes}
kinematics (see text). Estimates of the statistical accuracy are
based on an integrated luminosity of $200\,\mathrm{pb^{-1}}$.\hfill\mbox{}}
\label{fig:hermasy}
\end{center}
\end{figure}
Next we consider the corresponding double-spin asymmetry $A_{LL}$ in 
Fig.~\ref{fig:hermasy}.
As in Fig.~\ref{fig:asycomp} we study the relevance of the ``resolved''
photon contribution in the upper panel and the dependence on models for the
non-perturbative partonic structure of circularly polarized photons in the
lower panel of Fig.~\ref{fig:hermasy}.
Estimates of the statistical accuracy, Eq.~(\ref{eq:all-error}),
are based on the integrated luminosity of  ${\cal{L}}=200\;\mathrm{pb}^{-1}$
actually collected by {\sc Hermes} and the parameters as specified above.
Here, the ``resolved'' photon processes cause a much less pronounced shift
in the asymmetry than for {\sc Compass}, see Fig.~\ref{fig:asycomp}.
Also, there is almost no difference between the results obtained with the
two extreme models for the $\Delta f^{\gamma}$ densities.
This is because for the same transverse momentum $p_{T,c}$ the {\sc Hermes}
experiment is closer to the end of phase-space than {\sc Compass}, i.e., 
$x_{T,c}=2p_{T,c}/\sqrt{S}$ is closer to one. On average {\sc Hermes} probes
larger momentum fractions both in the nucleon and in the photon which explains
our results.
Our estimate of the statistical accuracy in Fig.~\ref{fig:hermasy} are such that 
only the region $1<p_{T,c}<1.5\;\mathrm{GeV}$ is potentially useful to further 
constrain the gluon polarization $\Delta g$. 
New experimental results from {\sc Hermes} for both $A_{LL}$ and the 
underlying unpolarized cross section will become available in the near future
\cite{ref:hermes-new} superseding the results of 
an earlier publication \cite{ref:hermes-2had}.
%
%%%%%%%%%%%%%%%
%%% FIGURE 8
%%%%%%%%%%%%%%%
\begin{figure}[tp]
\begin{center}
\vspace*{-.8cm}
\includegraphics[width=0.49\textwidth,clip=]{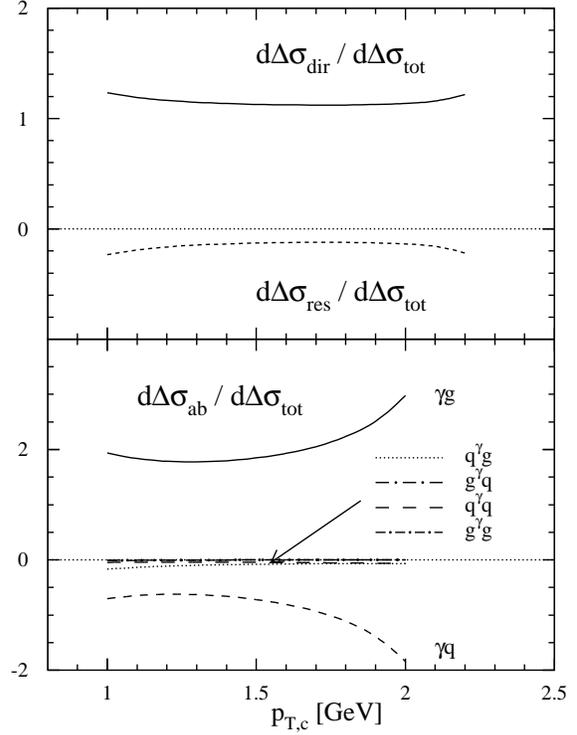}
\caption{\sf As in Fig.~\ref{fig:ratiocomppol} but now for \textsc{Hermes}
kinematics (see text).\hfill\mbox{}}
\label{fig:ratiohermpol}
\end{center}
\end{figure}

We finish this section with a detailed study
of the different contributions to the polarized photoproduction cross section.
The upper and lower panel of Fig.~\ref{fig:ratiohermpol} show the fractional
contributions of $d\Delta\sigma_{\mathrm{dir}}$ and  $d\Delta\sigma_{\mathrm{res}}$
and of the different partonic LO channels $a+b\to c+d$, respectively.
As for Fig.~\ref{fig:ratiocomppol} we use the maximal positive gluon polarization of GRSV
with $\Delta g=g$ at the input scale.
Again the choice is due to the fact that for our ``default'' gluon polarization,
GRSV ``standard'', the cross section develops a node at some $p_{T,c}$ such
that ratios are difficult to visualize.
As is already expected from Fig.~\ref{fig:hermasy}, the 
``resolved'' photon cross section is fairly small throughout,
though not completely negligible. Again it contributes with 
the opposite sign than the ``direct'' cross section, cf.~Fig.~\ref{fig:ratiocomppol}.
As for {\sc Compass} kinematics before, we observe a partial cancellation
of the two ``direct'' channels: photon-gluon fusion and QCD Compton scattering.
None of the four ``resolved'' contributions makes a significant contribution at LO.

%%%%%%%%%%%%%%%%%%%%%%%%%%%%%%%%%%%%%%%%%%%%%%%%%%%%%%%
\subsection{Prospects for a Future Polarized $ep$ Collider }
%%%%%%%%%%%%%%%%%%%%%%%%%%%%%%%%%%%%%%%%%%%%%%%%%%%%%%%
The most interesting option for a future experimental spin physics facility
is a first polarized lepton-proton collider such as the 
eRHIC project at BNL \cite{ref:erhic} currently under discussion. 
Here we consider the asymmetric collider option using the existing 
$250\,\mathrm{GeV}$ proton beam of RHIC and a new $10\,\mathrm{GeV}$ 
electron beam, i.e., a c.m.s.\ energy of $\sqrt{S}=100\,\mathrm{GeV}$.
The physics program of such a machine is based on the
extensive and highly successful exploration of unpolarized 
$ep$ collisions at the DESY-HERA collider. 
The determination of the gluon density 
from scaling violations in DIS down to very small $x$ and
establishing the concept of photonic parton densities in
photoproduction processes are some of the many physics highlights of HERA.
%
%%%%%%%%%%%%%%%
%%% FIGURE 9
%%%%%%%%%%%%%%%
\begin{figure}[tp]
\begin{center}
\vspace*{-.8cm}
\includegraphics[width=0.49\textwidth,clip=]{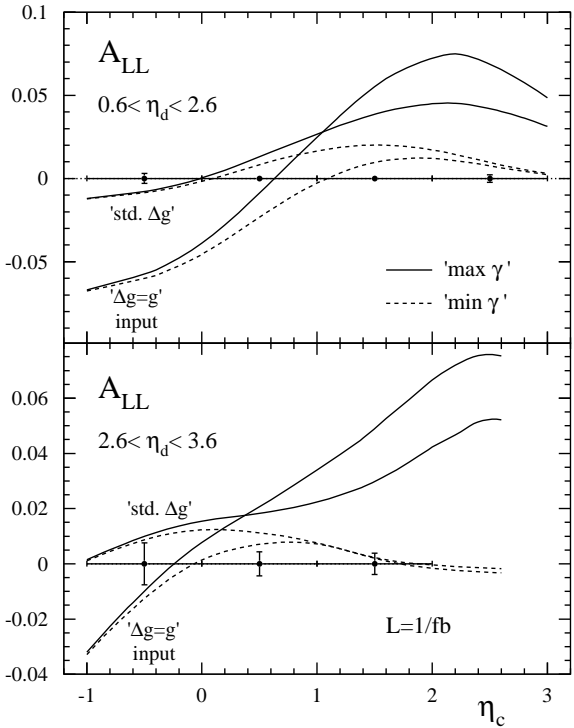}
\caption{\sf Double-spin asymmetry for photoproduction of two charged
hadrons in LO at $\sqrt{S}=100\,\mathrm{GeV}$. Results are given for
two different gluon polarizations of the nucleon and the two scenarios
for the $\Delta f^{\gamma}$ densities.
Both hadrons are required to have a transverse momentum of at least
$2\,\mathrm{GeV}$. The laboratory rapidity of one of the hadrons is integrated in
the range $0.6\leq \eta_d\leq 2.6$ and  $2.6\leq \eta_d\leq 3.6$ in the
upper and lower panel, respectively.
Estimates of the statistical accuracy refer to an integrated luminosity
of $1\,\mathrm{fb}^{-1}$.
\hfill\mbox{}}
\label{fig:asy-erhic}
\vspace*{-0.5cm}
\end{center}
\end{figure}

From similar precision studies of scaling violations in polarized DIS at eRHIC
it would be possible to access the gluon polarization 
down to $x\simeq 10^{-3}$, about one decade in $x$ lower than in $pp$ collisions at RHIC.
In photoproduction processes it is expected that ``resolved'' contributions
play a much more significant role at eRHIC than at fixed-target experiments
\cite{ref:polhera}.
This offers exciting prospects to learn about the $\Delta f^{\gamma}$ densities
as was already demonstrated in case of  single-inclusive pion 
photoproduction in Ref.~\cite{ref:pionerhic}.

Studies of hadron-pair production have the advantage that one has a better control
on the momentum fractions probed in the hadron and in the photon. For instance,
by demanding a ``trigger hadron'' in the proton's forward direction and 
scanning the other hadron's rapidity one can select kinematical
regions which are particularly sensitive to the non-perturbative structure of the
photon. This is demonstrated in Fig.~\ref{fig:asy-erhic} where we give estimates
for $A_{LL}$ at $\sqrt{S}=100\,\mathrm{GeV}$. The laboratory rapidity of the 
``trigger'' hadron is integrated in the range $0.6\leq \eta_d\leq 2.6$ and  
$2.6\leq \eta_d\leq 3.6$ in the upper and lower panel, respectively, and the 
rapidity of the other hadron is left unintegrated. 
For $\eta_c\lesssim 0$ our results are fairly independent of the choice
of a particular model for $\Delta f^{\gamma}$ while for $\eta_c\gtrsim 1$
the results strongly depend on $\Delta f^{\gamma}$. This is in particular
the case if the trigger hadron is detected more forward in the incoming proton's direction,
i.e., for $2.6\leq \eta_d\leq 3.6$ (lower panel of  Fig.~\ref{fig:asy-erhic}).
Note that in this subsection positive rapidities denote the proton direction,
in line with common conventions usually used at HERA.
Again this behavior can be understood by looking at the typical momentum
fractions $x_{\gamma}$ probed in Eq.~(\ref{eq:xsec}): for $\eta_{c,d}$ large and positive
one finds $x_{\gamma}\ll 1$ and for  $\eta_{c,d}$ large and negative $x_{\gamma}$
approaches one.

We also give estimates for the statistical accuracy in Fig.~\ref{fig:asy-erhic}
based on beam polarizations of ${\cal{P}}_{e,p}=0.7$
and an integrated luminosity of $1\,\mathrm{fb}^{-1}$. 
The latter is expected to be accumulated within only a few weeks of
running\linebreak eRHIC so that the statistical accuracy can be eventually much better than
the one shown in Fig.~\ref{fig:asy-erhic}. In addition, we demand a minimum
transverse momentum for both hadrons of $2\,\mathrm{GeV}$ and
$z_{c,d}>0.1$.
For the equivalent photon spectrum in Eq.~(\ref{eq:wwspectrum}) we use
similar parameters as the H1 and ZEUS experiments at HERA:
$Q_{\max}^2=0.5\,\mathrm{GeV}^2$ and the momentum fraction taken
by the photon is limited to $0.2\leq y \leq 0.85$. 
Note that the ``boost'' factor between the laboratory and the $ep$-c.m.s.\ 
frame is very similar for the asymmetric collider option for
eRHIC and for HERA. 

Keeping in mind that the polarized gluon distribution $\Delta g$ should
be known fairly well in the range relevant in
Fig.~\ref{fig:asy-erhic}, i.e.\ $x\gtrsim 0.01$, from RHIC by the time eRHIC would
start to operate, there are excellent prospects to study the so far unknown
parton content of circularly polarized photons.

%%%%%%%%%%%%%%%%%%%%%%%%%%%%%%%%%%%%%%%%%
\section{Conclusions}
%%%%%%%%%%%%%%%%%%%%%%%%%%%%%%%%%%%%%%%%%
In summary, we have presented a phenomenological study of 
spin-dependent photoproduction of hadron pairs at \linebreak c.m.s.\ energies
relevant for the {\sc Compass} and {\sc Hermes} experiments as well as
for a possible future polarized $ep$ collider.
So far, our studies are limited to the LO approximation of QCD but
will be amended to NLO accuracy in due time.

We have consistently included the ``direct''
and ``resolved'' photon contributions to the photoproduction
cross section. It turned out that the ``resolved'' part leads to a 
significant shift in the experimentally relevant double-spin asymmetries 
and has to be accounted for in future analyzes of data. 
Fixed-target experiments are, however, mainly sensitive to the perturbative
``pointlike'' part of the photon structure.
This simplifies attempts to extract the gluon polarization $\Delta g$
which is the main goal of {\sc Compass} and one of the goals of {\sc Hermes}.
The non-perturbative parton content of circularly polarized photons
can be pro\-bed in detail in photoproduction processes at
higher c.m.s.\ energies which hopefully will become available at
some point in the future. By that time we most likely have a good
knowledge of the spin structure of the nucleon, in particular, from
ongoing measurements at RHIC.

The double-spin asymmetries for both \textsc{Compass} and \textsc{Hermes}
show the expected sensitivity to the gluon polarization in the
nucleon. Keeping in mind the very sizable scale dependence at LO
all results for $A_{LL}$ have to be taken with a grain of salt unless
the applicability of perturbative methods for two-hadron production
at comparatively low transverse momenta and c.m.s.\ energies has
been thoroughly investigated and demonstrated.
This is best achieved by comparing the underlying
unpolarized cross sections for the production of hadron-pairs
with theoretical expectations.
If these checks are passed, the production of hadron pairs 
in lepton-nucleon collisions will be an interesting and
complementary tool to further constrain the 
polarized gluon density at momentum fractions 
of about $x=0.1\div 0.2$. 
If it turns out that data and theory do not match, 
these measurements will open up a window to study the effects
of all-order resummations, the relevance of higher-twist
corrections, and perhaps the transition to the non-perturbative
regime so far little explored and understood.

%%%%%%%%%%%%%%%%%%%%%%%%%%%%%%%%%%
\section*{Acknowledgements}
%%%%%%%%%%%%%%%%%%%%%%%%%%%%%%%%%%
We are grateful to E.C.\ Aschenauer, P.\ Liebing, and V.\ Mexner ({\sc Hermes}) and to
H.\ Fischer, S.\ Hedicke, F.-H.\ Heinsius and J.-M.\ Le Goff ({\sc Compass}) for valuable
discussions and information about the experimental details of their photoproduction
measurements. 
C.H.\ was supported by a grant of the ``Bayerische Elitef\"orderung''.
This work was supported in part by the ``Deutsche Forschungsgemeinschaft (DFG)''.
%
%%%%%%%%%%%%%%%%%%%%%%%%%%%%%%%%%%%%%%%%%%%%%%%%%%%%

%

\begin{thebibliography}{} %%%%%%%%%%%%%%%%%%%%%%%%%%
%%%%%%%%%%%%%%%%%%%%%%%%%%%%%%%%%%%%%%%%%%%%%%%%%%%%
%
\bibitem{ref:rith-review} See, for example: K.\ Rith, in proceedings of the
{\em 12th International Workshop on Deep Inelastic Scattering (DIS2004)},
Strbske Pleso, Slovakia, D.\ Bruncko et al.\ (eds.), 2004, p.\ 183.
%
\bibitem{ref:jaffe-manohar} See, for example: R.L.\ Jaffe and A.\ Manohar,
Nucl. Phys. {\bf B337} (1990) 509.
%
\bibitem{ref:rhic-review} See, for example: G.\ Bunce, N.\ Saito, J.\ Soffer, and
W.\ Vogelsang, Annu.\ Rev.\ Nucl.\ Part.\ Sci.\ {\bf 50}, 525 (2000);
C.\ Aidala et al., {\em Research Plan for Spin Physics at RHIC}, 2005, BNL report
BNL-73798-2005.
%
\bibitem{ref:rhic-unpol} {\sc Phenix} Collaboration, S.S.\ Adler {\em et al.},
Phys. Rev. Lett. {\bf 91}, 241803 (2003); {\em ibid.} {\bf 95}, 202001 (2005);
Phys. Rev. {\bf D71}, 071102(R) (2005);
{\sc Star} Collaboration, J.\ Adams {\em et al.}, Phys. Rev. Lett. {\bf 92}, 171801 (2004);
{\tt nucl-ex/0602011};
M.L.\ Miller, {\sc Star} Collaboration, {\tt hep-ex/0604001}, to appear in the proceedings
of the {\em Particles and Nuclei International Conference (PANIC 05)}, 
Santa Fe, New Mexico, Oct.\ 2005.
%
\bibitem{ref:rhic-pol} {\sc Phenix} Collaboration, S.S.\ Adler {\em et al.},
Phys. Rev. Lett. {\bf 93}, 202002 (2004); {\tt hep-ex/0602004};
K.\ Boyle, {\sc Phenix} Collaboration,  talk presented at the
{\em Particles and Nuclei International Conference (PANIC 05)}, 
Santa Fe, New Mexico, Oct.\ 2005, to appear in the proceedings;
J.\ Kiryluk, {\sc Star} Collaboration, {\tt hep-ex/0512040}, to appear in the proceedings
of the {\em Particles and Nuclei International Conference (PANIC05)}, Santa Fe, NM,
Oct.\ 2005.
%
\bibitem{ref:compass} {\sc Compass} Collaboration, G.\ Baum {\em et al.},
CERN/SPSLC 96-14 (1996).
%
\bibitem{ref:hermes} {\sc Hermes} Collaboration, {\em The {\sc Hermes} Physics
Program \& Plans for 2001-2006}, DESY-PRC, 2000.
%
\bibitem{ref:twohadron} A.\ Bravar, D.\ von Harrach, and A.\ Kotzinian,
Phys. Lett. {\bf B421}, 349 (1998);
%
\bibitem{ref:hermes-2had} {\sc Hermes} Collaboration, A.\ Airapetian {\em et al.},
Phys. Rev. Lett. {\bf 84}, 2584 (2000).
%
\bibitem{ref:smc-2had} Spin Muon Collaboration ({\sc Smc}), B.\ Adeva {\em et al.},
Phys. Rev. {\bf D70}, 012002 (2004).
%
\bibitem{ref:compass-2had} {\sc Compass} Collaboration, E.S.\ Ageev {\em et al.},
Phys. Lett. {\bf B633}, 25 (2006).
%
\bibitem{ref:pp-trouble} {\sc E706} Collaboration, L.\ Apanasevich {\em et al.},
Phys. Rev. Lett. {\bf 81}, 2642 (1998); P.\ Aurenche {\em et al.},
Eur. Phys. J. {\bf C9}, 107 (1999); {\em ibid.} {\bf 13}, 347 (2000);
U.\ Baur {\em et al.}, {\tt hep-ph/0005226};
C.\ Bourrely and J. Soffer, Eur. Phys. J. {\bf C36}, 371 (2004).
%
\bibitem{ref:upcoming} C.\ Hendlmeier {\em et al.}, work in progress.
%
\bibitem{ref:conto} J.J.\ Peralta, A.P.\ Contogouris, B.\ Kamal, and F.\ Lebessis, 
Phys. Rev. {\bf D49}, 3148 (1994);
G.\ Grispos, A.P.\ Contogouris, and G.\ Veropoulos, Phys. Rev. {\bf D62}, 014023 (2000).
%
\bibitem{ref:erhic} See {\tt http://www.bnl.gov/eic} for information concerning 
the eRHIC/EIC project, including the ``whitepaper'', BNL-report 68933,
Feb.\ 2002; A.\ Deshpande, R.\ Milner, R.\ Venugopalan, and W.\ Vogelsang,
Ann. Rev. Nucl. Part. Sci. {\bf 55}, 165 (2005).
%
\bibitem{ref:polhera} M.\ Stratmann and W.\ Vogelsang,
Z. Phys. {\bf C74}, 641 (1997).
%
\bibitem{ref:fact} S.B.\ Libby and G.\ Sterman, Phys.\ Rev.\ {\bf D18}, 3252 (1978);
R.K.\ Ellis, H.\ Georgi, M.\ Machacek, H.D.\ Politzer,
and G.G.\ Ross, Phys. Lett. {\bf 78B}, 281 (1978); Nucl. Phys. {\bf B152}, 285 (1979);
D.\ Amati, R.\ Petronzio, and G.\ Veneziano,
Nucl. Phys. {\bf B140}, 54 (1980); Nucl. Phys. {\bf B146}, 29 (1978);
G.\ Curci, W.\ Furmanski, and R.\ Petronzio, Nucl.\ Phys.\ {\bf B175}, 27 (1980);
J.C.\ Collins, D.E.\ Soper, and G.\ Sterman, Phys.\ Lett.\ {\bf B134}, 263 (1984);
Nucl.\ Phys.\ {\bf B261}, 104 (1985);
J.C.\ Collins, Nucl.\ Phys.\ {\bf B394}, 169 (1993).
%
\bibitem{ref:ww} D.\ de Florian and S.\ Frixione, Phys. Lett. {\bf B457},
236 (1999).
%
\bibitem{ref:kkp} B.A.\ Kniehl, G.\ Kramer, and B.\ P\"{o}tter, Nucl. Phys.
{\bf B582}, 514 (2000).
%
\bibitem{ref:cteq} {\sc Cteq} Collaboration, 
J.\ Pumplin {\it et al.}, JHEP 0207, 012 (2002).
%
\bibitem{ref:grvphoton} M.\ Gl\"{u}ck, E.\ Reya, and A.\ Vogt,
Phys. Rev. {\bf D46}, 1973 (1992).
%
\bibitem{ref:grsv}  M.\ Gl\"{u}ck, E.\ Reya, M.\ Stratmann, and
W.\ Vogelsang, Phys. Rev. {\bf D63}, 094005 (2001).
%
\bibitem{ref:polphoton} M.\ Gl\"{u}ck and W.\ Vogelsang, Z. Phys. {\bf C55} (1992) 353;
Z. Phys. {\bf C57} (1993) 309; M.\ Gl\"{u}ck, M.\ Stratmann, and W.\
Vogelsang, Phys. Lett. {\bf B337} (1994) 373;
M. Stratmann and W. Vogelsang, Phys. Lett. \textbf{B386} (1996) 370.
%
\bibitem{ref:onehadron} B.\ J\"{a}ger, M.\ Stratmann, and W.\ Vogelsang,
Eur. Phys. J. {\bf C44} (2005) 533.
%
\bibitem{ref:pionpp} B.\ J\"{a}ger, A.\ Sch\"{a}fer, 
M.\ Stratmann, and W.\ Vogelsang, Phys. Rev. {\bf D67} (2003) 054005; 
%
\bibitem{ref:pionerhic} B.\ J\"{a}ger, M.\ Stratmann, and W.\ Vogelsang, 
Phys. Rev. {\bf D68} (2003) 114018.
% 
\bibitem{ref:resum} See, e.g., D.\ de Florian and W.\ Vogelsang,
Phys. Rev. {\bf D71} (2005) 114004; Phys. Rev. {\bf D72} (2005) 014014;
and references therein.
%
\bibitem{ref:hermes-new} E.C.\ Aschenauer and P. Liebing, private communications.
%
\end{thebibliography}
\end{document}